\documentclass[twocolumn,journal]{IEEEtran}
\IEEEoverridecommandlockouts
\usepackage[dvips]{graphicx}
\usepackage{subfigure}
\usepackage[english]{babel}
\usepackage{algorithm}
\usepackage{algorithmic}
\usepackage{amsfonts}
\usepackage{booktabs}
\usepackage{multirow}
\usepackage[numbers,sort&compress]{natbib}
\usepackage{amsmath}
\usepackage{amssymb}
\usepackage{array}
\usepackage{url}
\usepackage{caption}

\usepackage[colorlinks,
linkcolor=black,
anchorcolor=black,
citecolor=blue]{hyperref}

\begin{document}

\title{Rethinking Blockchains in the Internet of Things Era from a Wireless Communication Perspective}

\def\showauthor{} %

\ifx\showauthor\undefined
\else

\author{
\IEEEauthorblockN{Hongxin Wei, Wei Feng, Yunfei Chen, Cheng-Xiang Wang, and Ning Ge}\\
\thanks{Hongxin Wei, Wei Feng (corresponding author), and Ning Ge are with Tsinghua University; Yunfei Chen is with the University of Warwick; Cheng-Xiang Wang is with Purple Mountain Laboratories and Southeast University.}
}

\fi

\maketitle
\begin{abstract}

Due to the rapid development of Internet of Things (IoT), a massive number of devices are connected to the Internet.
For these distributed devices in IoT networks, how to ensure their security and privacy becomes a significant challenge.
The blockchain technology provides a promising solution to protect the data integrity, provenance, privacy, and consistency for IoT networks.
In blockchains, communication is a prerequisite for participants, which are distributed in the system, to reach consensus.
However, in IoT networks, most of the devices communicate through wireless links, which are not always reliable. Hence, the communication reliability of IoT devices influences the system security.
In this article, we rethink the roles of communication and computing in blockchains by accounting for communication reliability. We analyze the tradeoff between communication reliability and computing power in blockchain security, and present a lower bound to the computing power that is needed to conduct an attack with a given communication reliability. Simulation results show that adversarial nodes can succeed in tampering a block with less computing power by hindering the propagation of blocks from other nodes.

\end{abstract}

\IEEEpeerreviewmaketitle

\section{Introduction}

The blockchain technology emerges as a distributed crypto-based ledger system, which runs in a decentralized mode and does not need a central controller.
Being open, transparent, traceable, and tamper-resistant, the blockchain technology provides a solution for trusted value exchange among the nodes regardless of whether a participator is a human or a machine. It does not require participators in the system to have any prior knowledge before trading, or to rely on any third-party authority such as a bank.
The blockchain technology is the key technology for Bitcoin, which was invented by Satoshi Nakamoto in 2008 \cite{Nakamoto2008Bitcoin}. Later, smart contracts are introduced into blockchains. Smart contracts are executable programs, which are stored in blockchains and run on blockchains to execute the agreement coded in the programs. As smart contracts become prevalent, the blockchain technology has found its applications in such areas as finance, banking, intellectual property, and logistics.

Along with the development of Internet of Things (IoT), more and more devices are connected to the Internet, such as intelligent vehicles \cite{Zhang2019Artificial}, devices for smart homes, home security, and health care. For these distributed devices in IoT networks, how to ensure their security and privacy becomes a significant challenge \cite{dai2019blockchain}. The blockchain technology provides a promising solution to protecting the data integrity, provenance, privacy, and consistency for IoT networks \cite{Zhang2019Edge}. Furthermore, by applying the blockchain technology to the exchange of services or data between IoT devices, the value of the information in IoT networks could be activated and amplified. Smart contracts will greatly stimulate the efficiency of information exchange in IoT networks. This will, in turn, trigger an increase of applications of IoT, such as software updates, sharing of service and property (e.g. house sharing, bicycle sharing, and car sharing), and smart supply chain with auto payment.

For most machine-to-machine devices and wearable devices in IoT networks, data are conveyed through wireless communications \cite{Zhang2019Deep}. In the future, there will be a lot of blockchain nodes communicating through wireless links.
However, wireless communication is not always reliable, and hence messages broadcast in wireless blockchain networks may be lost. Therefore, the outage probability of the wireless links in a blockchain system may influence the performance of the system. Nakamoto analyzed the influence of computing on the security of Bitcoin, and concluded that, if attacks attempt to replace a block, they have to own more than 50 percent of the total computing power in a blockchain network, which is usually referred to as ``51\% attack'' \cite{Nakamoto2008Bitcoin}. However, in Nakamoto's analysis, the communication reliability of the network is not considered.

Several researchers have discussed the influence of communication on the performance of blockchains.
 Decker and Wattenhofer showed that information propagation speed could influence the performance of blockchain systems \cite{Decker2013Information}.
Danzi \emph{et al.} analyzed the effect of communication quality on blockchain synchronization for IoT networks \cite{danzi2018analysis}.
Sun \emph{et al.} analyzed the transaction throughput and communication throughput in blockchain-enabled wireless IoT networks \cite{8668426}. In their analysis, node geographical distribution and transaction arrival rate are modeled as Poisson point processes.
Kim analyzed the impact of mobility on blockchain performance in vehicular ad-hoc networks \cite{8720159}.
In \cite{Heilman2015Eclipse} and \cite{Nayak2016Stubborn}, the authors presented an ``eclipse attack'' algorithm, in which adversarial nodes can earn more rewards by manipulating the communication in the networks. To conduct an eclipse attack, attackers need to monopolize all the connections of the victim node, thus other nodes cannot receive blocks from the victim node and the attackers will be more likely to win the mining competition.

In this article, we focus on the influence of communication reliability on the security of blockchains. The tradeoff between communication reliability and computing power for blockchains is analyzed. Simulation results show that by affecting the communication reliability of the network, adversarial nodes can use less computing power to tamper a block, which has been confirmed by the blockchain network.
The main contributions of this article include:
\begin{itemize}
     \item The blockchain technology is rethought from a communication perspective. The roles of communication and computing in blockchains are clarified.
    \item The tradeoff between communication reliability and computing power in blockchain security is analyzed. We also present a lower bound of computing power that is needed to conduct an attack with a given communication reliability.
\end{itemize}

The rest of the article is organized as follows. In Section II, we briefly introduce the fundamentals of blockchain technology and the role of communications in blockchains. In Section III, the tradeoff between communication reliability and computing power is analyzed. In Section IV, simulation results are presented, and some open issues are discussed. Finally, conclusions are given in Section V.

\section{The role of communication in blockchains}

\begin{figure}[tbp]
\centering
\includegraphics[width=3.5in]{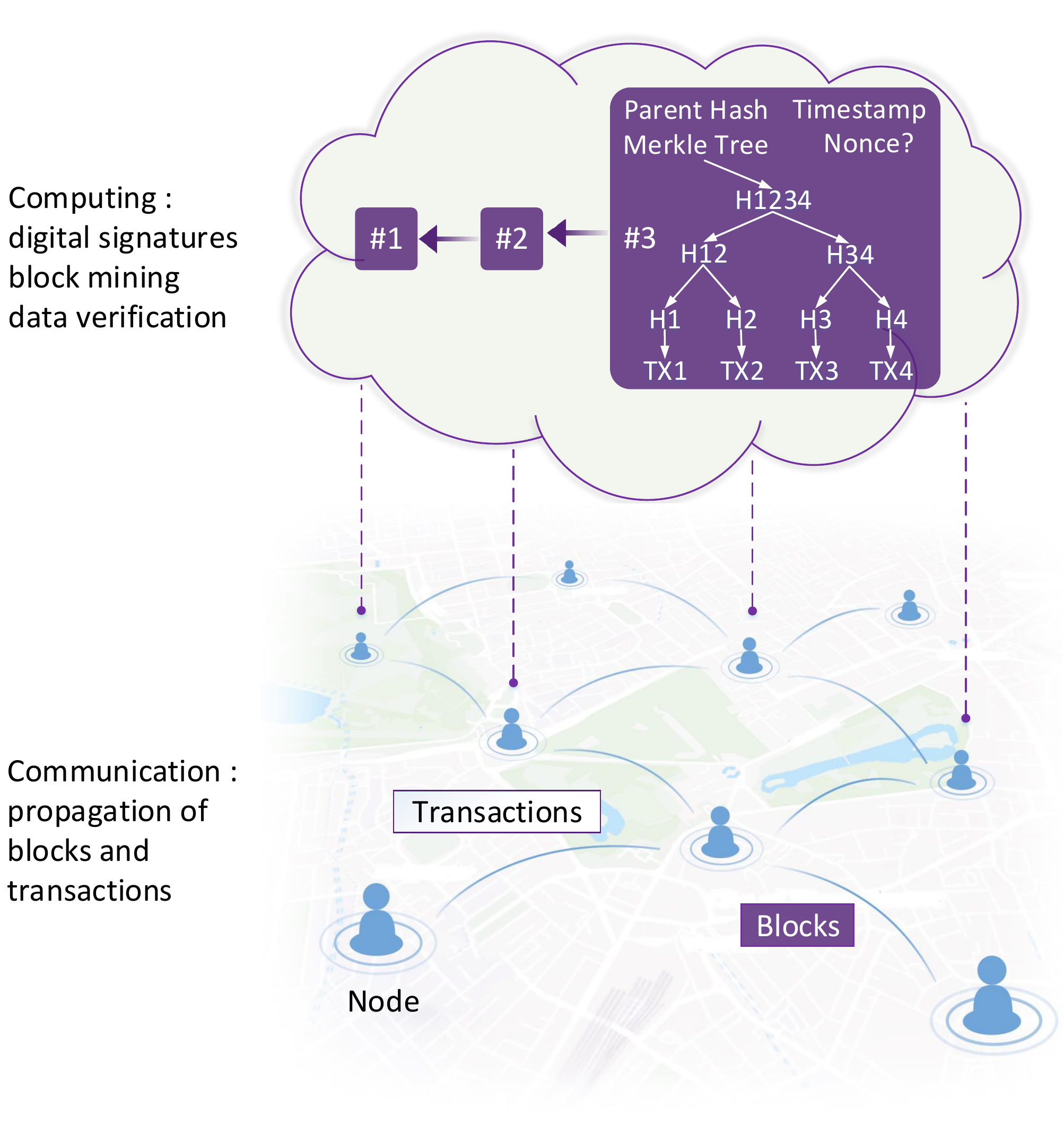}
\caption{Illustration of the roles of communication and computing in blockchains.}
\label{fig-blockchainlayers}
\end{figure}

The blockchain technology is rooted in the synergy of communication and computing.
Communication and computing jointly make the distributed ledgers reliable and consistent.
In Fig. \ref{fig-blockchainlayers}, we explain the roles of communication and computing in blockchains.

In a blockchain, data are stored in blocks, and blocks are organized in a chain. The chained blocks are stored in distributed nodes, and each node could keep a complete replica of the whole chain.
In Fig. \ref{fig-blockchainlayers}, Block \#3 shows the data structure of Bitcoin.
In each block, there are two parts. The first part is the block head. Some key data fields in the block head of Bitcoin are listed in the figure.
The other part is the body of the block. In a blockchain system, data are written in transactions, and transactions are stored in the body of the block.
The block head contains four critical fields: the parent hash, the Merkle tree hash \cite{Merkle1987digital}, the nonce, and the timestamp.
The parent hash is the hash of the previous block. It guarantees that the blocks are linked together in series. The Merkle tree hash is the root value of the Merkle tree, which represents all the transactions in the current block. In a Merkle tree, the value of a leaf node is the hash of the transaction data, and the value of a non-leaf node is the hash of its children's values. In Bitcoin, leaf nodes keep the hash of transactions. The value of the Merkle tree root could be used to verify whether all the data in leaf nodes of the tree have been changed. If the value of one node is changed, the value of its parent will change. As a result, the value of the Merkle tree root will change, and the hash of the block will change. For example, in Block \#3 in Fig. \ref{fig-blockchainlayers}, ``TX1'' and ``TX2'' are transaction data, ``H1'' and ``H2'' are the hashes of ``TX1'' and ``TX2'', ``H12'' is the hash of ``H1H2'', the Merkle tree root ``H1234'' is the hash of ``H12H34''. If ``TX1'' is tampered, then ``H1'', ``H12'', ``H1234'' and the hash of the block will change.
The nonce is a random number, which is used to determine who generates the current block.
The timestamp is used to record the time when the block is created. It provides a proof of existence at a certain time in the right order.

In a typical blockchain network, all nodes are equal and connected with each other in a peer-to-peer (P2P) mode. Every node exchanges information, including transactions and blocks, with its neighboring nodes. This is different from the conventional Client-Server (CS) or Browser-Server (BS) network structure, where all the data are transmitted to a server first and then the server processes and forwards them to the destinations.
Some nodes access the network through wired links, and the others through wireless links.
Any device running blockchain programs can join the blockchain network, such as a computer, a smartphone, a pair of smart glasses, a smart car, a smart refrigerator, or a router.

\begin{figure*}[tbp]
\centering
\includegraphics[width=7.4in]{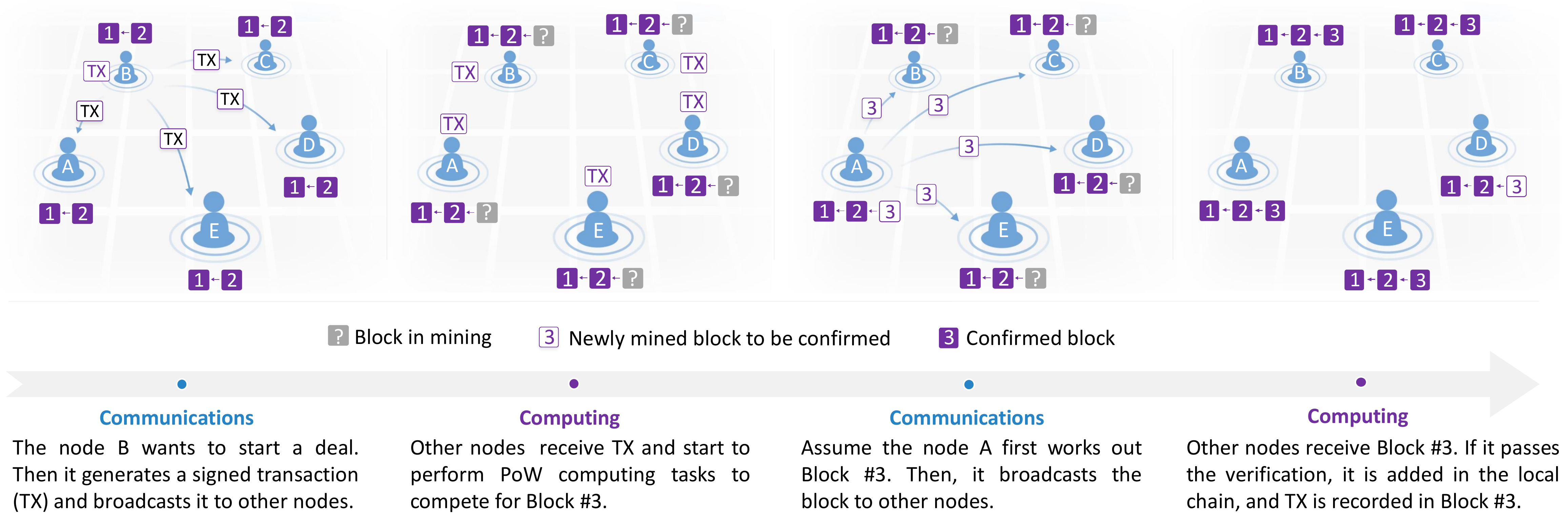}
\caption{Transaction procedures of a blockchain using PoW consensus.}
\label{fig-blockchaintx}
\end{figure*}

In Fig. \ref{fig-blockchaintx}, we present a case study of Bitcoin to demonstrate how a blockchain works \cite{Nakamoto2008Bitcoin}.
In the system, there are five nodes: A, B, C, D, and E. Each node has a copy of the entire blockchain. Initially, the blockchain contains two blocks: Block \#1 and Block \#2. The key information that is needed to perform a transaction on Bitcoin network is contained in the blocks. In a typical blockchain system, the transaction process consists of four main steps:

\begin{enumerate}
 \item Transmission of transactions. Assume that a new transaction is to commence at node B. The transaction is digitally signed by the payer via encrypting the transaction with its private encryption key. The digital signature can ensure that no one else can create fake transactions on behalf of the payer. Denote the digitally signed transaction as TX. Then, TX is broadcast from node B to other nodes.

 \item Computing to generate a new block. Upon receiving TX, other nodes can place TX into a transaction pool to form a new block. Assume all nodes pack TX into a new block as a candidate for Block \#3, and start performing calculation tasks for winning the competition for Block \#3. The nodes participating in the calculation tasks are called miners. The operation of performing calculation tasks is called mining.
For Bitcoin, miners attempt to find a target nonce, which makes the hash of the block head less than a target value. The target value is an indicator of the difficulty of mining a new block. With a smaller target value, it is more difficult to find a target nonce. Once a target nonce is found, the block is generated successfully, and the miner will be rewarded with some coins.

 \item Transmission of the newly mined block. Assume the node A first finds a target nonce and works out Block \#3. Then, it broadcasts the block to other nodes as soon as possible to get the block confirmed by others.

 \item Verifying the validity of the received block. After a node receives Block \#3, it attempts to verify the validity of the received block by checking whether the data format of the block is correct, whether all transactions in the block have been signed correctly and their payers can afford them, and whether the nonce and timestamp of the block are legal. If the block passes the verification, it will be stored in the local database. Through these processes, the transaction is recorded in the blockchain across the network.

\end{enumerate}

In a blockchain, computing mainly occurs in the consensus mechanisms, including producing digital signatures in Step 1, computing hash in Step 2, and verifying data in Step 4. Consensus mechanisms are used to guarantee that the ledgers distributed at different nodes will reach consistency in the long term. Otherwise, if everyone possesses a different version, the ledgers could not function well. Take Bitcoin as an example, the consensus mechanism that searches for a target nonce like solving puzzles is called proof-of-work (PoW). PoW is currently used in the two most popular blockchain systems: Bitcoin and Ethereum \cite{Wood2014Ethereum}. In PoW consensus mechanisms, if an attacker attempts to change the $i$th block, it has to conduct all the PoW calculation tasks from the $i$th block to the latest block. These tasks need a great deal of computing power in Bitcoin and Ethereum networks, as the computing power for them is already very high. Hence, there is barely an incentive for attackers to tamper data in old blocks.
Notably, in PoW consensus mechanisms, searching for a target nonce is not very efficient and consumes a lot of energy. Hence, some other consensus mechanisms are explored, such as Paxos, practical byzantine fault tolerant (PBFT),  proof-of-stake (PoS), delegated proof-of-stake (DPoS), and proof-of-activity (PoA) \cite{Tschorsch2016Bitcoin}. 

{The role of communication in blockchains is to enable the highly distributed data in all the nodes to finally reach a consensus and increase the total consensus information of the system.}
When some data need to be stored in a blockchain, miners need to perform computing tasks to search for proper nonces, and the total amount of information in all nodes increases. But the information in consistency does not increase simultaneously, as the information at distributed nodes has not reached a consensus. Via communications between the nodes, all the nodes can reach consistency, and the total amount of consensus information increases.

\section{Tradeoff between communication reliability and computing power}

From the procedure described in Fig. \ref{fig-blockchaintx}, it can be seen that transactions and blocks need to be broadcast among all nodes.
If the block is not spread to other nodes in Step 3, it is uncertain that Block \#3 will be successfully accepted by other nodes, and attackers may have more chances to tamper the data.
Therefore, the security of the system depends not only on the computing power, but also on the communication reliability.

Consider a simplified scenario, where there exists one hostile node while other nodes are honest. The hostile node tries to produce blocks that contain fake transactions to steal digital assets in the blockchain system. For example, the attacker may create two transactions for different destination nodes, in order to implement a double-spend attack \cite{Nakamoto2008Bitcoin}. Assume the probability that the hostile node works out a new block earlier than the honest nodes is $q_w, 0\leq q_w \leq 1$, and the probability that honest nodes succeed is $p_w=1-q_w$. During the communication process, the block may be lost or go wrong because of channel fading, noise, and interference in wireless communications, network congestion, and node breakdown. Assume the probability that honest nodes succeed in broadcasting the block in a round is $q_c, 0\leq q_c \leq 1$, where $q_c$ measures the communication reliability of honest nodes.

Denote $q$ as the probability that the hostile node wins in a round of competition for generating a new block.
In a round, $q$ could be calculated as $\frac{q_w}{q_w+p_wq_c}$.
If the hostile node attempts to tamper the data in a block, assume the block is $z$th ahead of the latest block in the blockchain, then it has to tamper the last $z$ blocks.
Assume $P(z)$ is the probability that the hostile node could finally catch up with other nodes and replace these blocks. Denote $Q=\frac{q_w}{p_wq_c}$. According to the result in \cite{Nakamoto2008Bitcoin}, $P(z)$ is one when $Q\geq 1$, and $Q^z$ when $Q < 1$.

For scenarios with multiple hostile nodes, if all the hostile nodes are cooperative, they can be regarded as a single hostile node, whose total computing power is the sum of all their computing power. The influence on the communication reliability of the honest node is the combination of all their influence. If all the hostile nodes are non-cooperative, each hostile node attacks independently. Thus, for each hostile node, the other hostile nodes can be viewed as honest nodes along with the actual honest nodes during its attacking process.

\begin{figure}[tbp]
   \centering
\includegraphics[width=3.4in]{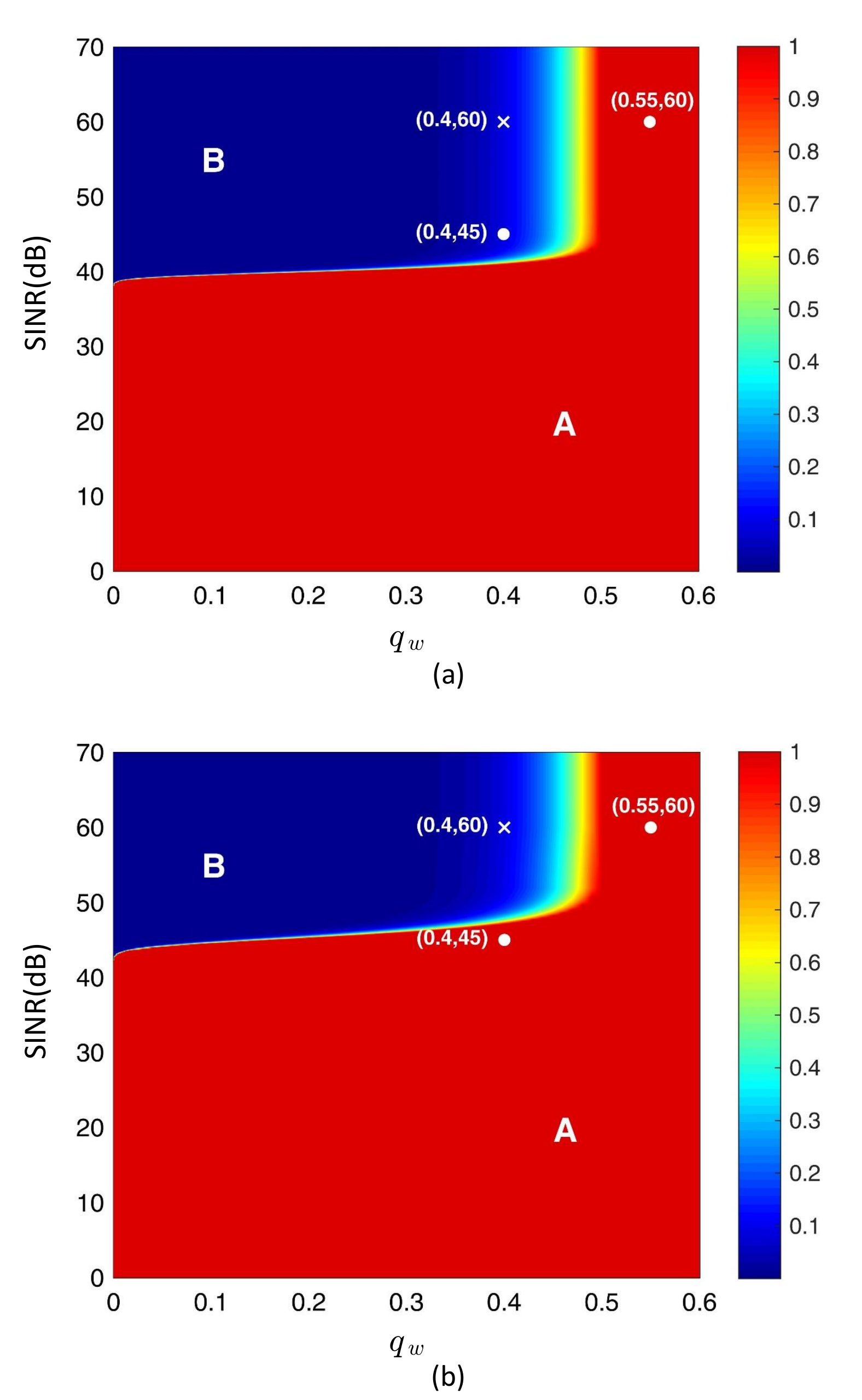}
	\caption{The probability that the attacker can succeed with six blocks behind: a) up to three retransmission attempts; b) up to six retransmission attempts.}
	\label{fig-tradeoff}
\end{figure}

In wireless communications, communication reliability mainly depends on the transportation strategy and bit error rate (BER). For a given modulation and coding scheme, BER is mainly decided by signal to interference plus noise ratio (SINR). Increasing SINR can decrease BER and increase communication reliability. Hence, we use SINR as a metric of communication reliability in the following analysis.
Figure. \ref{fig-tradeoff} shows the mean of $P(z)$ over 1000 trials with different pairs of ($q_w$, SINR). In the simulation, QPSK modulation and Rayleigh fading channel with a scale parameter of 0.5 are used. $z$ is set to 6 and the block size is set to 8 Mbits. Each block is divided into 1000 packets. If a packet is lost or any error occurs in the packet transmission, the honest node can retransmit it until the number of retries has reached a predetermined maximum retry count. In Fig. \ref{fig-tradeoff}(a) and Fig. \ref{fig-tradeoff}(b), the honest node is allowed to make up to three and six retransmission attempts, respectively.

 As shown in Fig. \ref{fig-tradeoff}(a), when $Q \geq 1$ as in Region A, the hostile node could catch up sooner or later. When $Q < 1$  as in Region B, the probability that the hostile node could succeed in attacking follows $Q^z$ with $z$. In the figure, the points $(0.55,60)$ and $(0.4,45)$ are located in Region A. The point $(0.4,60)$ is located in Region B.
When SINR is bigger than 50 dB, the variation of SINR has little influence on $P(z)$. This is because, when SINR is high, $q_c$ approximately equals 1, and $Q$ approximately equals $\frac{q_w}{1-q_w}$. In this case, increasing SINR can hardly promote $q_c$ and $Q$. Thus, $P(z)$ mainly depends on $q_w$. This is similar to the case in the conventional analysis, which assumes perfect communications. If $q_w > 0.5$ such as the point $(0.55,60)$, the attacker will succeed eventually. If $q_w < 0.5$ such as the point $(0.4,60)$, it is not certain that the attacker will succeed.
However, when communication reliability is considered, attackers may need less computing power, and the ``51\% attack'' does not hold anymore. Even if the computing power is less than 50 percent, the attacker can still conduct an attack by hindering the propagation of the honest nodes. For example, the attacker can generate artificial noise to lower the SINR at the honest nodes. The probability of a successful attack can be further increased by increasing the power of artificial noise. When SINR is lower than 40dB, $q_c$ is close to zero. In this case, the attacker can conduct a successful attack with little computing power.

Comparing Fig. \ref{fig-tradeoff}(a) and Fig. \ref{fig-tradeoff}(b), it is clear that the point $(0.4,45)$ is located in Region A when the maximum retry count is 3, and in Region B when the maximum retry count is 6. This indicates that increasing the number of retransmission attempts can help the honest node lower the risk of being compromised by attackers.

\section{Simulations and analysis}
In this section, we perform numerical simulations to demonstrate the influence of communications in a blockchain system.
In the simulations, there exists one hostile node and one honest node, and the honest node stands for all other honest nodes.
In each round of the mining competition, the hostile node and the honest node perform computing tasks to compete for generating the new block. When the honest node succeeds, it broadcasts the block to other nodes through wireless communication channels. The wireless channels are modeled as Rayleigh fading with parameter 0.5, and gray-coded QPSK modulation is used. Each block contains 8 Mbits and is divided into 1000 packets to transmit. If a packet is lost, the node is allowed to make up to three retransmission attempts. When the honest node receives a block from the hostile node, which has a bigger block number than the longest chain in its buffer, it will accept the block from the hostile node. In this state, the attack of the hostile node is regarded as successful.

\begin{figure}[tbp]
\centering
\includegraphics[width=3.5in]{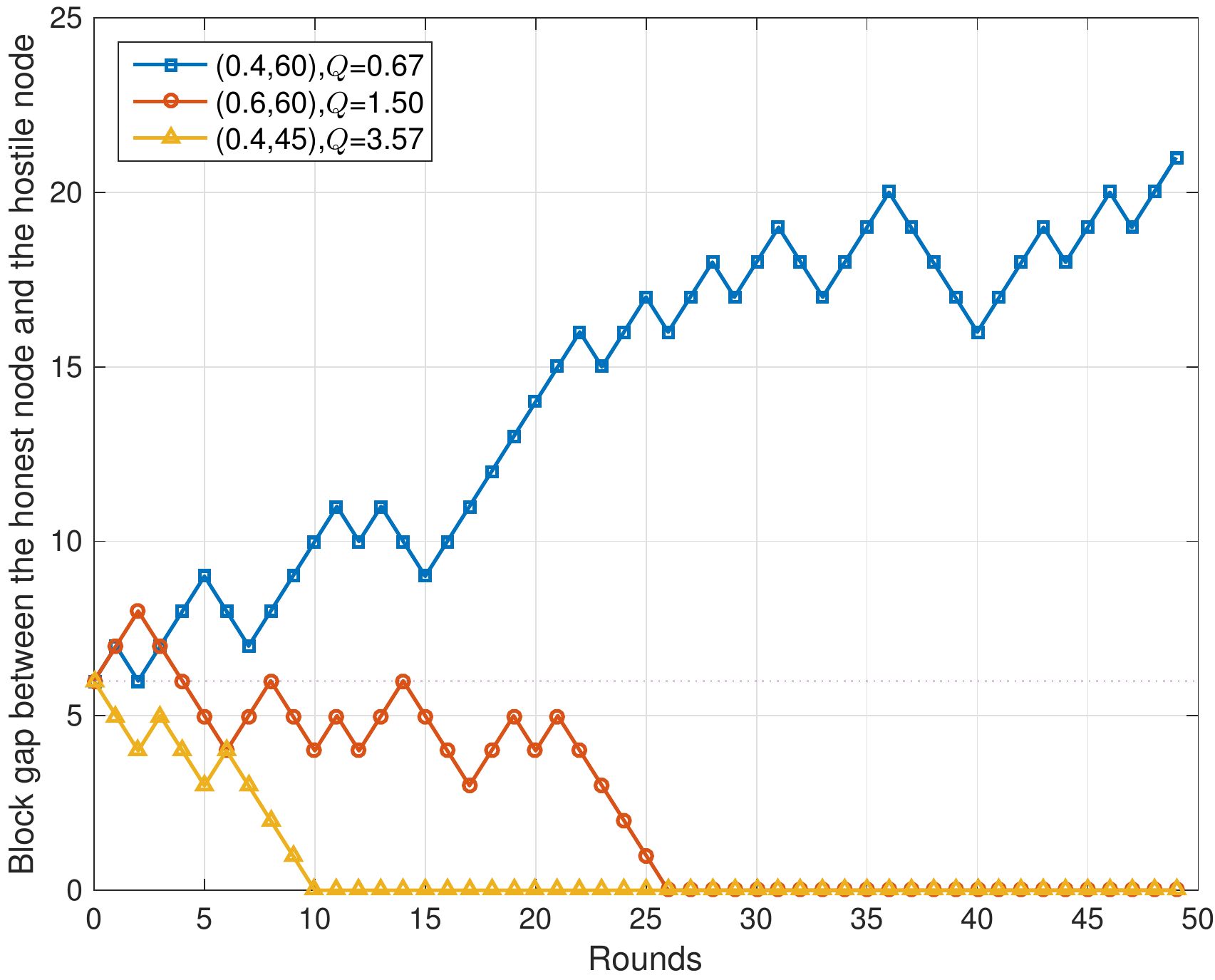}
\caption{Block gap between the hostile node and the honest node in each round. }
\label{fig:gap:zgap}
\end{figure}

Figure. \ref{fig:gap:zgap} shows the block gap between the honest node and the hostile node at each round. Different pairs of ($q_w$, SINR) are simulated. The initial value of $z$ is set to 6. When $Q \geq 1$ such as the points $(0.6,60)$ and $(0.4,45)$, the hostile node could eventually catch up, thus the gap gradually converges to zero.
When $Q < 1$ such as point $(0.4,60)$, the hostile node has the chance to succeed, but not 100 percent. And the gap expands gradually in the long term.

\begin{figure}[tbp]
\centering
\includegraphics[width=3.5in]{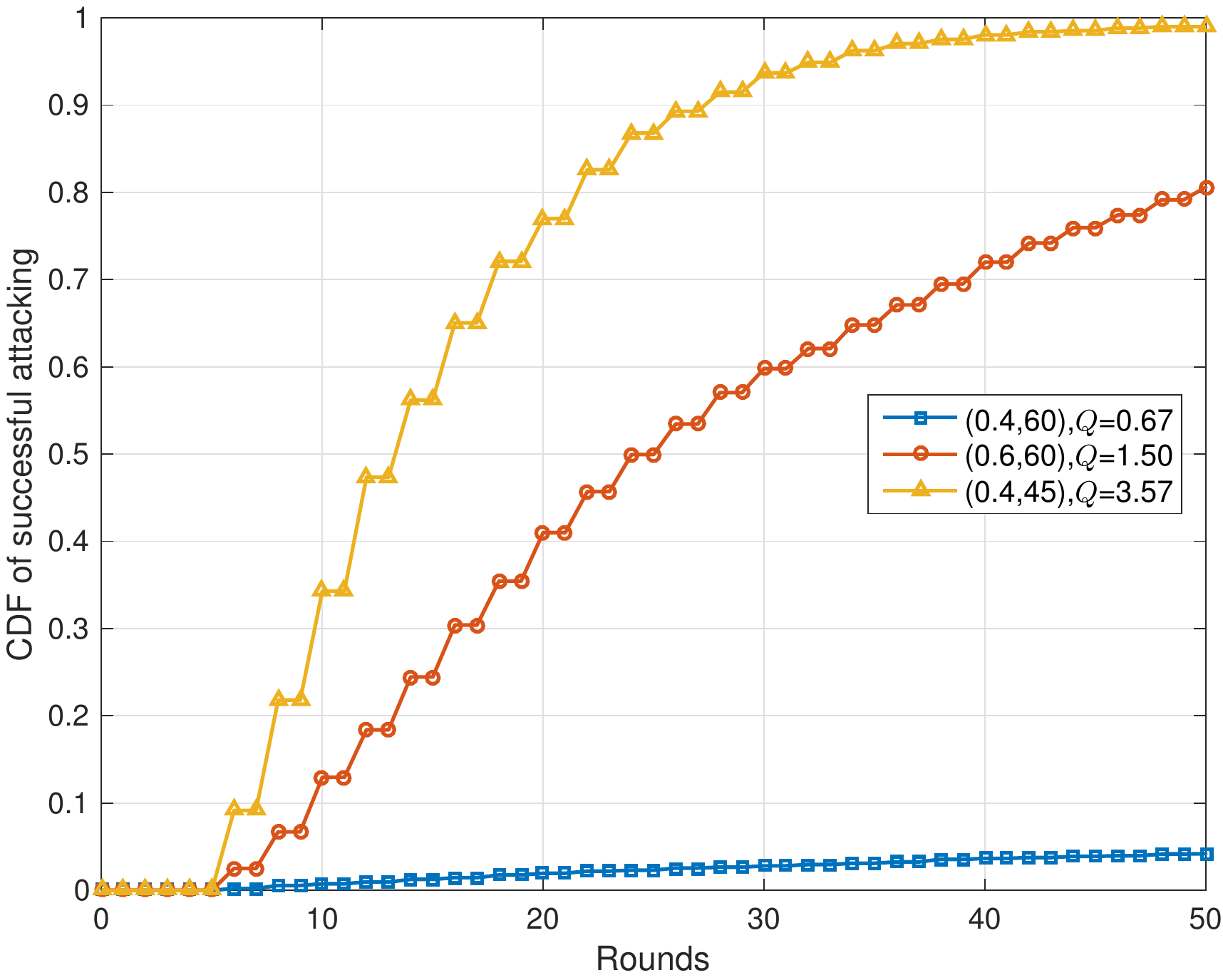}
\caption{CDF for different ($q_w$, SINR) in each round.                    }
\label{fig:gap:gapcdf}
\end{figure}

In Fig. \ref{fig:gap:gapcdf}, the average cumulative distribution function (CDF) of a successful attack over 1000 trials is shown. The initial value of $z$ is set to 6. This figure shows the hostile node's attacking process for three pairs of ($q_w$, SINR) at different block numbers.
When $Q \geq 1$, the CDF of a successful attack finally reaches 100 percent. And with larger $Q$, the attacking process will be faster.
When $Q < 1$, the CDF of successful attack increases gradually, but the speed is slower.
Notably, the CDF in each odd round is the same as that in the previous round. Let us assume that the hostile node can succeed in each round. Then, it can catch up with the honest node in the sixth round. If it fails in one of the first six rounds, after the sixth round, it still needs to win two blocks to catch up with the honest node. Hence, it cannot succeed in the seventh round, and the probability that the hostile node succeeds within seven rounds is the same as that within six rounds.

\begin{figure}[tbp]
\centering
\includegraphics[width=3.5in]{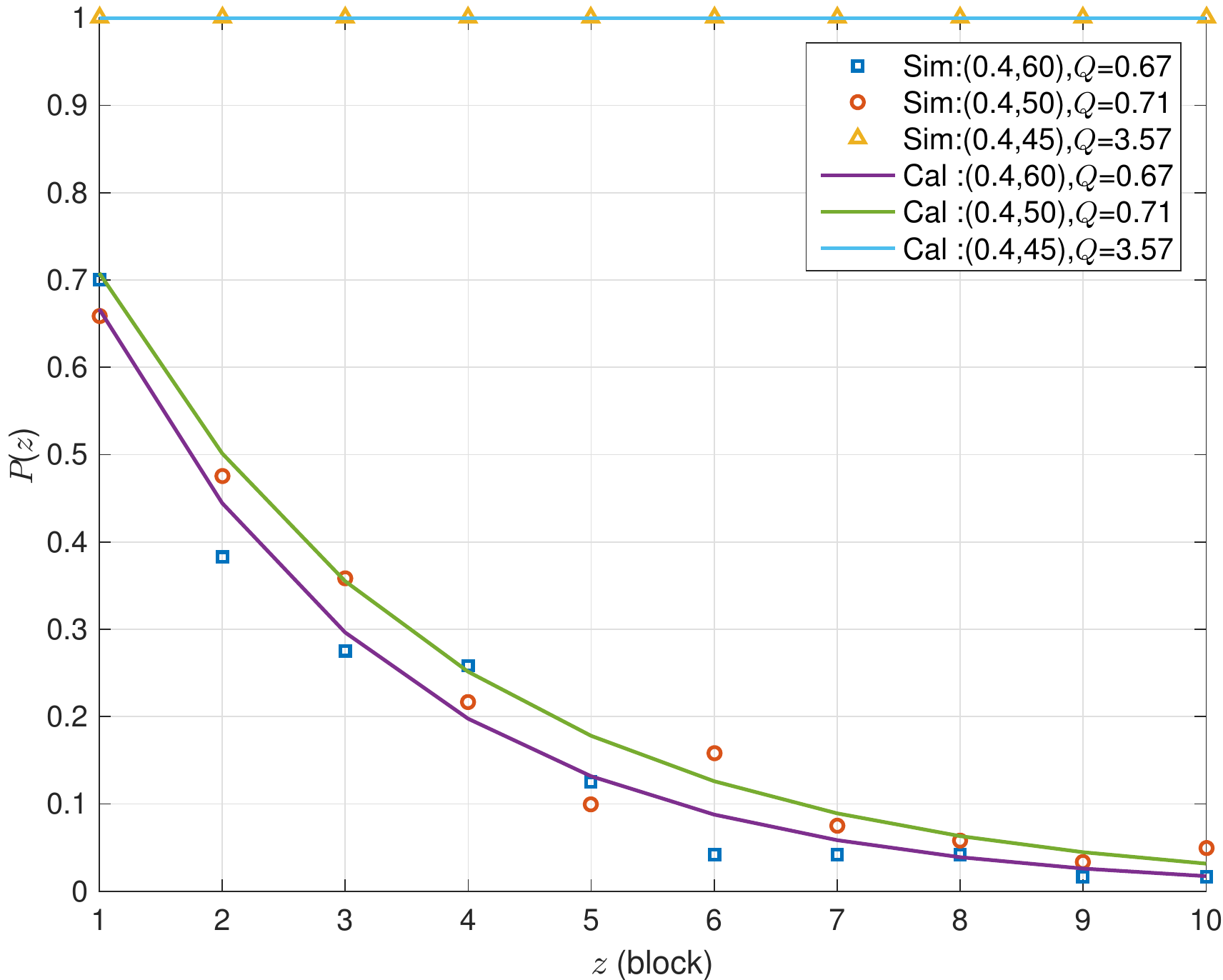}
\caption{The probability that the attacker succeeds when $z$ varies. }
\label{fig-PzVSz}
\end{figure}

In Fig. \ref{fig-PzVSz}, we simulate the influence of $z$ with different pairs of ($q_w$, SINR).
 The probability of a successful attack is calculated as the probability that the hostile node has caught up with the honest node after a fixed number of blocks in order to take place of some confirmed transactions.
In the simulation, the number is set to 1000, which is long enough to cover most of the cases.
In the figure, the average success probability over 1000 trials is shown with markers and the theoretical value of $P(z)$ is shown in curves. As is shown, the simulation results perfectly match the calculations.
When $Q \geq 1$ such as the point $(0.4,45)$, the hostile node could catch up finally, thus the probability is 100 percent at different values of $z$.
when $Q < 1$ such as the points $(0.4,50)$ and $(0.4,60)$, the probability that the hostile node could succeed converges to zero exponentially.
Comparing these two curves, we can see that with smaller $Q$, $P(z)$ drops faster as $z$ increases.

\section{Open research issues}

From the simulation and analysis, we have shown the tradeoff between communication reliability and computing power in blockchains. However, there are still some open research issues that need further study.

\begin{enumerate}
 \item{Transportation strategies.}

 In a practical blockchain system, nodes may employ some transportation control strategies such as automatic repeat-request (ARQ). The influence of communications on blockchains with different transportation strategies needs more in-depth research.

 \item{Network properties.}

 In this article, communication reliability is characterized by the average probability that an honest miner succeeds in transmitting its newly mined block ahead of the hostile nodes. However, some network properties also affect communication reliability. For example, the bandwidth and the transmission delay of honest nodes affect the time, which is needed to propagate a block to their neighbors. The network size and the number of each node's neighbors affect the total time needed to propagate a block to all nodes. These parameters will affect whether a hostile node can broadcast its newly mined block before other nodes. Therefore, more details about network settings should be taken into account. 

\item{Mining strategies.}

 In a blockchain, miners can decide when to broadcast their newly mined blocks and whether to accept and relay the blocks they received or not. Under different mining algorithms, the impacts of communications are different and necessary to be analyzed separately.

\end{enumerate}

At present, the performance of blockchains remains to be a setback for the popularization. In blockchains, the broadcast of P2P networks is implemented by sending data to each node in unicast mode. This is quite inefficient.
In a blockchain system, information dissemination has the following characteristics:
 \begin{enumerate}
 \item Broadcast.

 In blockchains, transactions and blocks should be broadcasted to all nodes.
 \item Massive data.

 When the number of nodes is large, the total amount of data that need to be spread through the P2P network are very huge.
 
 \item Global decentralization.

 For public blockchains, nodes are distributed around the world. The route path of block dissemination is very long, resulting in a very long time to diffuse a block.
 \end{enumerate}

If the broadcast property of wireless communications could be incorporated into the data dissemination in blockchains, the transmission efficiency will be enhanced.
Furthermore, satellite communications are featured with the advantages of wide-range coverage and broadcast, which are suited to be applied in blockchain systems to broadcast data. Thus, the transmission efficiency of blockchains can be improved further.

On the other hand, blockchains can also be used in wireless communication systems. With the features of decentralization, integrity, traceable, tamper-resistant, programmable smart contract, authority, and anonymity, the blockchain technology has been used in resource management, access management, and system configuration.
In a conventional system, these functions mainly rely on centralized management organizations. However, these organizations may abuse user data, leak user privacy, and be vulnerable to single point failure and denial of service (DoS) attack. Moreover, many management organizations are very inefficient, which causes a waste of resources. The blockchain technology provides a promising solution to these problems.

Meanwhile, blockchain-based networks, especially IoT networks, will produce massive data. It is hard to process and analyze these data manually. Artificial intelligence (AI) technologies can be used to handle this problem. Through data analysis and mining, AI can help to manage the system more flexibly and intelligently. Authors in \cite{8726067} have integrated blockchain and AI to build a secure intelligent network for 5G beyond. In the future, more work about the integration of blockchain and AI is needed to make the networks more secure, intelligent, and efficient.

\section{Conclusion}
In this article, we have clarified the roles of communication and computing in blockchains. Furthermore, we have analyzed the influence of communication reliability and computing power on blockchain security for a PoW blockchain system, as well as the tradeoff between them. Simulation results have shown that by hindering the propagation of legal blocks from other nodes, attackers could use less computing power to replace blocks, which have been confirmed by the blockchain networks. Therefore, the researchers of blockchains need to pay more attention to the attack against the communication networks. In future research, models of transportation strategies, network properties, and mining strategies need further study. Furthermore, AI technology can be integrated into the blockchain-empowered IoT network to make it more intelligent and efficient.

\ifx\showauthor\undefined
\else
\section*{Acknowledgement}
This work was supported in part by the National Key R\&D Program of China (Grant No. 2018YFA0701601), the National Natural Science
Foundation of China (Grant No. 61922049, 61771286, 61941104, 61701457), the Nantong Technology Program (Grant No. JC2019115),
and the Beijing Innovation Center for Future Chip.
\fi

\ifx\showauthor\undefined
\else

\section*{Biographies}
\noindent
HONGXIN WEI is a postdoctoral research fellow with the Department of Electronic Engineering, Tsinghua University. His research interests include blockchain, energy-efficient wireless networks, and coordinated satellite-terrestrial networks.
\\

\noindent
WEI FENG [S'06, M'10, SM'19] is an associate professor with the Department of Electronic Engineering, Tsinghua University. His research interests include maritime broadband communication networks, large-scale distributed antenna systems, and coordinated satellite-terrestrial networks. He serves as the Assistant to the
Editor-in-Chief of China Communications, an Editor of IEEE TCCN, and an Associate Editor of IEEE Access.
\\

\noindent
YUNFEI CHEN [S'02, M'06, SM'10] is an associate professor with the School of Engineering, University of Warwick, U.K. His research interests include wireless communications, cognitive radios, wireless relaying, and energy harvesting. 
\\

\noindent
CHENG-XIANG WANG [S'01, M'05, SM'08, F'17] is a professor with Purple Mountain Laboratories and the School of Information Science and Engineering, Southeast University, Nanjing, China.
His research interests include wireless channel measurements/modeling, B5G wireless communication networks. He is an Executive Editorial Committee (EEC) Member for IEEE Transactions on Wireless Communications. He is a fellow of the IEEE and IET.
\\

\noindent
NING GE is a professor with the Department of Electronic Engineering, Tsinghua University. His research interests include communication ASIC design, short-range wireless communications, and wireless communications.

\fi

%
%
%
%
%

\end{document}